# Influence of plasmon resonances and symmetry effects on second harmonic generation in WS$_2$-plasmonic hybrid metasurfaces

Florian Spreyer[1], Claudia Ruppert[2], Philip Georgi[1], and Thomas Zentgraf[1]

[1]*Department of Physics, Paderborn University, Warburger Straße 100, 33098 Paderborn, Germany*

[2]*Experimentelle Physik 2, Technische Universität Dortmund, Otto-Hahn-Straße 4a, 44227 Dortmund, Germany*

**Abstract:**

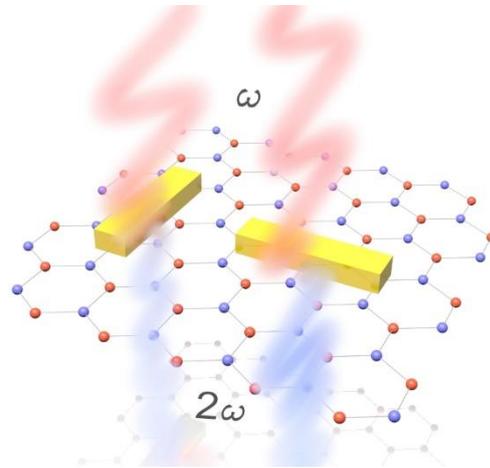

The nonlinear process of second harmonic generation (SHG) in monolayer (1L) transition metal dichalcogenides (TMD), like WS$_2$, strongly depends on the polarization state of the excitation light. Combining plasmonic nanostructures with 1L-WS$_2$ by transferring it onto a plasmonic nanoantenna array, a hybrid metasurface is realized impacting the polarization dependency of its SHG. Here, we investigate how plasmonic dipole resonances affect the process of SHG in plasmonic-TMD hybrid metasurfaces by nonlinear spectroscopy. We show, that the polarization dependency is affected by the lattice structure of plasmonic nanoantenna arrays as well as by the relative orientation between the 1L-WS$_2$ and the individual plasmonic nanoantennas. In addition, such hybrid metasurfaces show SHG in polarization states, where SHG is usually forbidden for either 1L-WS$_2$ or plasmonic nanoantennas. By comparing the SHG in these channels with the SHG generated by the hybrid metasurface components, we detect an enhancement of the SHG signal by a factor of more than 40. Meanwhile, an attenuation of the SHG signal in usually allowed polarization states is observed. Our study provides valuable insight into hybrid systems where symmetries strongly affecting the SHG and open up the possibility for tailoring the SHG in 1L-WS$_2$ for future applications.

In the past years, two-dimensional layered materials gained interest in diverse fields of physics. At the latest, transition metal dichalcogenides (TMDs) are prominent materials and their properties are investigated intensely. TMDs are layered materials and formed by one metal atom M and two chalcogen atoms X, forming the generic formula $MX_2$.[1] They are usually present in bulk material and consist of stacked atomic layers connected by van der Waals forces, making it possible to scale them down to the size of a monolayer with atomic thickness (1L). As a result, the band gap in these semiconductors changes from an indirect to a direct band gap.[2]

The band gap energy varies for different TMD materials and, for $WS_2$, can reach up to 2 eV[3], making them interesting for optics in a wide wavelength regime. In addition, the comparatively high absorbance of light exceeding 15% for atomic layers of less than 1 nm thickness[4] is a great advantage, allowing an easy characterization of 1L-TMD with optical methods like absorbance/transmittance measurements and photoluminescence or Raman-spectroscopy[5-8]. Hence, the unique properties of 1L-TMD open up the possibility for implementations in a wide range of promising applications like transistors, photodetectors, and diodes.[9-11]

Another field of application, where 1L-TMD draws attention, is the field of plasmonics. Its applications use the same wavelength range from the visible up to near-infrared (NIR) while offering additional properties like strong near-field interactions and large scattering cross-sections.[12-13] The wide range of applications and well-known physics of tunable plasmonic nanostructures provide a great fundament for implementing them in different types of hybrid metasurfaces.[14-16] The combination of 1L-TMDs with plasmonics nanostructures can provide strong light-matter interaction while maintaining very small scales for light manipulation and the possibility of using well-known fabrication methods like electron beam lithography for plasmonic metasurfaces. Although the fabrication process of large-area single-crystalline flakes of 1L-TMD is still challenging, recent progress in monolayer fabrication (e.g. mechanical exfoliation, chemical vapour deposition) is paving the way towards large scale applications,[17-20] making it easier for transferring 1L-TMDs onto plasmonic metasurfaces. The resulting hybrid metasurfaces consist of plasmonic nanoantenna arrays and 1L-TMDs and are recently in focus of intense research.[21-23]

Particular interest is being paid to the field of tailored nonlinear optics with nanoscale materials. As 1L-TMDs possess a hexagonal lattice structure with two different atoms M and X, the inversion symmetry is broken and therefore allows second harmonic generation (SHG). For 1L-TMDs, the properties of SHG are quite well-known and utilized for various purposes,[24-26] exhibiting great potential for a wide range of applications, e.g. beam steering, lensing, phase modulation and nanocavity-enhanced SHG.[27-30] In comparison, plasmonic nanoantennas can also exhibit nonlinear harmonic generation and its nonlinear order is related to the local symmetry of the nanostructure.[31] For single nanorods with a two-fold (C2) rotational symmetry, the inversion symmetry is not broken and, therefore, SHG is forbidden in the dipole approximation, while for a three-fold rotational symmetry C3, a significant SHG signal can be measured. Although coupling effects in hybrid metasurfaces consisting of plasmonic nanoantenna arrays and 1L-TMD are reported for SHG,[27, 32-34] the impact of the individual symmetries of each part of such hybrid metasurfaces on the SHG process is still largely unexplored.

Here, we experimentally investigate the impact of symmetry effects on the SHG process of hybrid metasurfaces consisting of a monolayer of $WS_2$ and a plasmonic metasurface made of Au nanorod antennas. In particular, we show how the polarization-dependent SHG signal of hybrid metasurfaces is determined by the 1L-TMD material properties and how it is influenced by the localized plasmonic resonances. By varying the relative orientation angle between the individual plasmonic nanorods and the 1L-TMD, the SHG signal arising from hybrid metasurfaces can be suppressed although it is allowed by the 1L-TMD. Furthermore, the impact of different lattice structures of plasmonic nanoantenna arrays is studied and how a present/absent broken inversion symmetry of the global lattice influences the SHG in hybrid metasurfaces. As a result, we show that plasmonic nanoantenna arrays open up channels for SHG into polarization states, that are forbidden by either 1L-$WS_2$ and the bare plasmonic nanoantennas resulting in an enhancement of SHG by a factor of more than 40 in these channels, depending on the different symmetries of the lattice structures. Our work demonstrates that different symmetries strongly alter the SHG in hybrid metasurfaces, which is of great interest for the design of hybrid nonlinear systems.

For our hybrid metasurfaces, a monolayer of $WS_2$ (1L-$WS_2$) is transferred on top of the plasmonic nanoantenna arrays that are placed on a fused quartz substrate (for details see Supplementary Material). To characterize the impact of the relative rotation angle between the hexagonal lattice structure of the 1L-$WS_2$ and the long axis of the gold nanorod, the fabricated hybrid metasurfaces are divided into four distinct areas, where each area consists of a plasmonic nanorods array where the nanorods are rotated by a specific angle $\alpha$ with respect to the $WS_2$ lattice (see Figure 1A). To study the impact of different lattice structures for the plasmonic nanoantenna arrays on the SHG signal, we fabricated three distinct samples, wherein each sample´s plasmonic nanoantenna array has a different lattice structure, with either a square, hexagonal or a degenerated hexagonal lattice structure (Figures 1B-D).

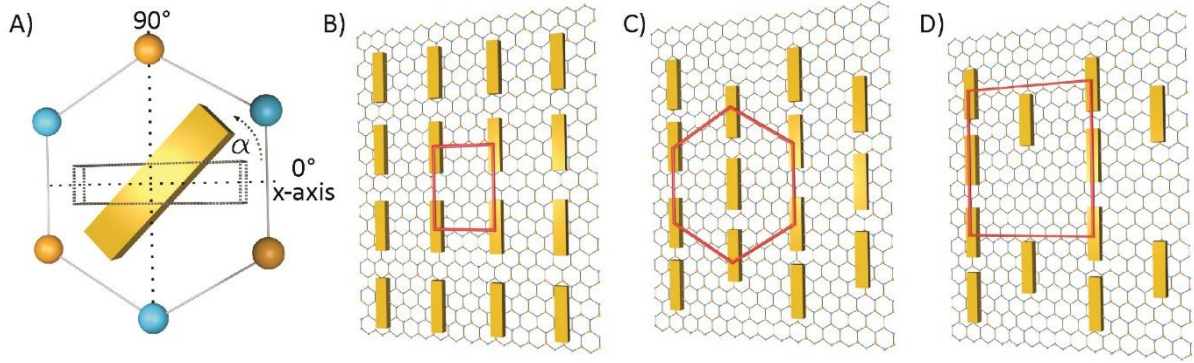

**Figure 1: Hybrid metasurface for second-harmonic generation. (A)** Relative rotation angle between plasmonic nanoantenna and the hexagonal lattice structure of WS$_2$ of 45-90° in 15° steps. The horizontal dotted line indicates the angle of 0° with respect to the hexagonal lattice structure. The blue and orange spheres represent the W and S atoms arranged in a hexagonal lattice. **(B-D)** Schematic illustration of the nanoantennas arranged in a square (B), hexagonal (C), and degenerated hexagonal (D) lattice structure. The red lines represent the unit cells for each lattice structure. The 1L-WS$_2$ layer depicted in these images lays on top of the nanoantenna arrays.

As already mentioned, for plasmonic nanorod antennas, the process of SHG is forbidden due to the given inversion symmetry. However, it is possible to introduce a broken inversion symmetry with the help of the lattice structure. This is possible if there is a coupling between neighboring nanoantennas. Such coupling can be either near-field coupling for small distances or radiative coupling.[35-36] Apparently, the square and hexagonal lattice structures possess inversion symmetry. Therefore, we introduce a degenerated hexagonal lattice structure, wherein the unit cell of the hexagonal lattice structure lacks the centered nanorod, while the non-horizontal next neighbor unit cells still contain the centered nanorod, creating a new unit cell with broken inversion symmetry (Figure 1D).

## Results & discussions

### Linear optical characterization

For our study, we fabricated six distinct metasurfaces where each lattice structure is fabricated twice. One of each is covered by the 1L-WS$_2$ to measure the SHG signal from hybrid metasurface and the second is the bare plasmonic metasurface as a reference. The nanorods are designed to be resonant at around 1230 nm, which equals half of the band gap energy of ~2 eV (≙615 nm) for 1L-WS$_2$. In this way, the plasmon resonances can resonantly couple to the fundamental wavelength, where a strong SHG signal from the 1L-WS$_2$ is expected. The individual nanorods are all of the same dimensions with a length of 300 nm and a width of 100 nm. For more details regarding the fabrication process and dimensions of the lattice structures see the Methods section and Supplementary Material. Scanning electron microscopy (SEM) images of the plasmonic nanoantenna arrays with different lattice structures are shown in Figures 2A-C.

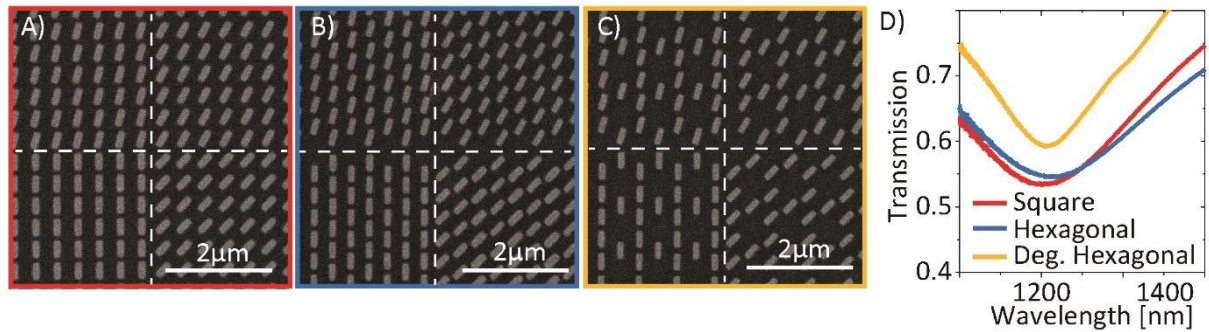

**Figure 2: Characterization of plasmonic nanoantenna arrays. (A-C)** Scanning electron microscopy images of the square (A), hexagonal (B), and degenerated hexagonal (C) lattice structure. Each metasurface contains four distinct areas, whereas within each area the plasmonic nanorods are rotated in steps of 15°. **(D)** Transmission spectra of the plasmonic nanoantenna arrays with square, hexagonal, and degenerated hexagonal lattice structure, measured with unpolarized light. The resonance wavelength was tuned to 1230 nm to match the resonance wavelength of the two-photon absorption process of 1L-WS$_2$.

For the optical characterization, we first measure the transmission spectrum of each fabricated metasurface with a Fourier-transformation-infrared spectrometer and unpolarized light (Figure 2D). As intended, all of the metasurfaces show a pronounced plasmon resonance in the transmission spectrum at around 1230 nm, although the transmission dip slightly varies in width, depth, and position. Previous studies have shown that this effect originates from the different types of lattice structures.[37] Note that the transmission spectra are taken from metasurfaces, containing all antenna rotations from 45-90° and not only for the individual rotation angles. After the characterization, we transferred a monolayer of WS$_2$ onto the plasmonic nanoantenna arrays and further measure the photoluminescence (PL) signal of the WS$_2$ by direct excitation with a wavelength of 532 nm. White-light images of the hybrid metasurfaces can be seen in Figures 3A-C. The transmitted PL light is filtered by a bandpass filter and analyzed via a spectrometer. The PL-images of the measured samples can be found in the Supplementary Material. Figure 3D shows the results, obtained from the WS$_2$ transferred onto the metasurface. We observe strong PL light, peaking at 613 nm with a linewidth of 10 nm for all samples, confirming that the transferred layers of WS$_2$ are of monolayer thickness. Comparing the PL-emission wavelength to the resonances of the plasmonic nanoantennas, we can confirm an overlap of the intended resonances.

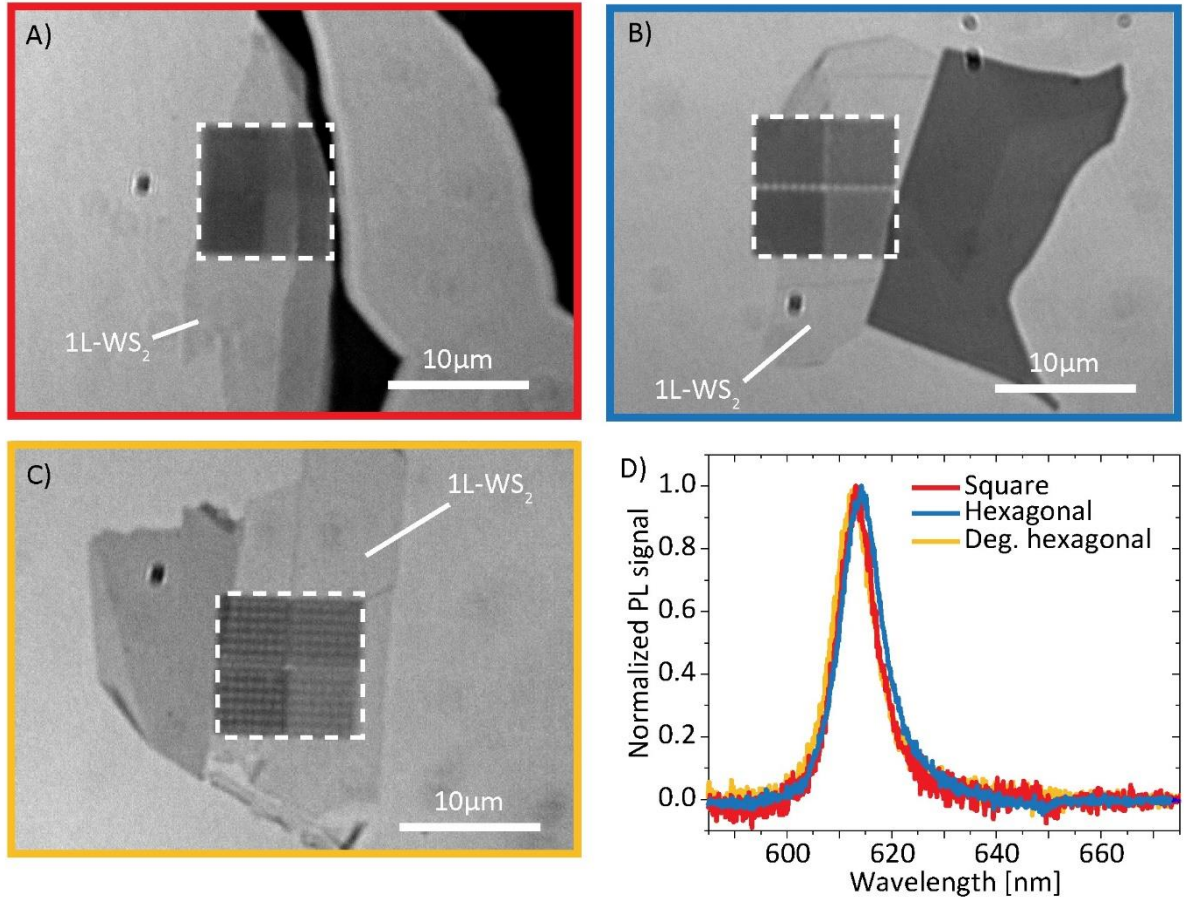

**Figure 3: Photoluminescence measurements. (A-C)** Optical microscopy images of the plasmonic nanoantenna array with the square (A), hexagonal (B), and degenerated hexagonal (C) lattice structure, covered with a 1L-WS$_2$ as indicated. The white dotted line marks the hybrid metasurface. **(D)** Normalized photoluminescence spectra of the three monolayers shown in (A-C) showing a clear peak at 613 nm indicating a band gap energy of ~2 eV.

## Nonlinear optical characterization

After the linear optical characterization, we analyzed the nonlinear optical properties of the hybrid metasurfaces. In a first step, we investigated the SHG signal strength of the 1L-WS$_2$ and bare plasmonic nanoantenna array independently from each other for different excitation wavelengths. For this, we measured the SHG signal with a home-build microscopy setup, depicted in Figure 4A. As fundamental pump light, we used laser pulses at wavelengths of 1200-1290 nm varied in 10 nm steps and 60 fs pulse length, emitted by an optical parametric amplifier system at a 1 MHz repetition rate. The polarization of the pump laser beam was set to a linear state by a combination of a linear polarizer and a half-wave plate. Subsequently, the excitation laser beam was filtered by a long pass (LP) and focused by a microscope objective down to a spot size of ~2 μm in diameter (FWHM). The beam illuminated the nanorods/1L-WS$_2$ on the backside of the sample, so that the beam passed the substrate first, hit the plasmonic nanoantennas subsequently and, in case of the hybrid metasurface, excited the 1L-WS$_2$ afterwards. The generated SHG signal was collected by another microscope objective and filtered by a short pass filter (SP) and another linear polarizer. Note, that the output linear polarizer was set to be parallel to the input polarization state. Concluding, the SHG signal was measured by a spectrometer. For the characterization of the SHG of the plasmonic nanoantenna array, we focus on the plasmonic nanoantennas arranged in the degenerated hexagonal lattice. Since the other samples contain plasmonic nanoantenna arrays without broken inversion symmetry, we only expect a strong SHG signal for the degenerated hexagonal lattice. The results of the measured SHG arising from the 1L-WS$_2$ and solely plasmonic nanoantenna array are shown in Figure 4B-C.

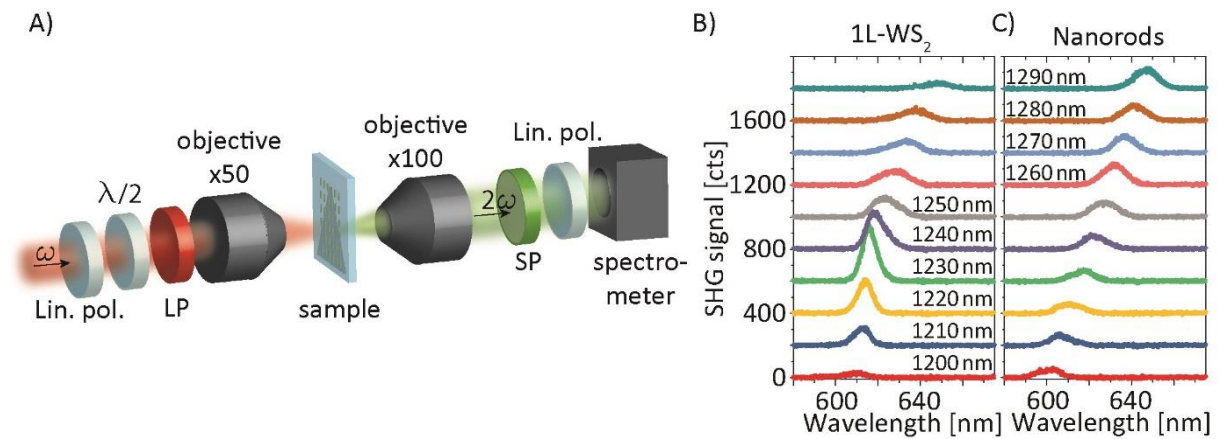

Figure 4: Wavelength-dependent SHG signal. **(A)** Schematic illustration of the setup for SHG measurement. The fundamental infrared laser beam first was linearly polarized, filtered by a long pass (LP) and then focused by a microscope objective (x50/NA 0.42) onto the backside of the sample. The input polarization was set by a linear polarizer (Lin. Pol.) and a half-wave plate. The SHG signal was collected by another microscope objective (x100/NA 0.8), filtered by a short pass (SP) and another linear polarizer, and measured by a spectrometer. **(B)** Measured SHG signal of the 1L-WS$_2$ for different excitation wavelengths. **(C)** Measured SHG signal of solely plasmonic nanoantenna arrays arranged for the degenerated hexagonal lattice structure and different excitation wavelengths. The input polarization was set to be alongside the nanorods long axis. Note that each spectrum is labeled with its excitation wavelength inside B) or C) and shifted upwards for better visibility.

In Figure 4B, it is shown, that the SHG signal arising from the 1L-WS$_2$ becomes strongest when excited with 1230 nm wavelengths. This fits well with the PL-measurement results, exhibiting a band gap energy of ~2 eV corresponding to 613 nm. Note, that for shorter excitation wavelengths, the nonlinear signal is not located at half the excitation wavelengths, but rather converges against 613 nm, hinting that the dominant nonlinear process is two-photon luminescence (TPL).[38-39] By measuring the SHG signal of the plasmonic nanoantenna array (Figure 4C), the strongest nonlinear signal cannot be easily identified, since the SHG intensity for most excitation wavelengths is nearly the same. This is due to the fact, that the plasmonic resonance is quite broad leading to a broad wavelength range for SHG. In comparison to the 1L-WS$_2$, the SHG wavelengths are always located at half the excitation wavelengths. In conclusion, we were using 1230 nm as excitation wavelength for all upcoming polarization-dependent measurements.

## Effects of symmetry on nonlinear harmonic generation

To investigate the SHG for the hybrid metasurface and its symmetry dependency, we measured the polarization-dependent SHG, to determine the influence of the coupling effects between the nanoantennas and the 1L-WS$_2$. During the transfer process of the 1L-WS$_2$, the orientation of the monolayer cannot be precisely set to the plasmonic nanorods' long axis, as illustrated in Figure 1A. Small variations from the ideal alignment are unpreventable and therefore, the relative orientation between the 1L-WS$_2$ and the nanorods long axis can slightly deviate from the intended angles. In the following discussion, we chose 0° to be the angle with respect to the 1L-WS$_2$ hexagonal lattice structure as depicted in Figure 1A and call it the x-axis. To investigate the impact of the different rotation angles of the plasmonic nanoantennas, we first determined the relative angle between the long axis of the nanoantenna and the fixed x-axis. Therefore, the nonlinear signals from the 1L-WS$_2$ and the plasmonic nanorods for different polarization angles were measured while the output linear polarizer is set to be parallel to the incident linear polarization of the fundamental beam at all times. This is crucial for investigating the impact of the different symmetries on the SHG signal since the SHG signal of solely 1L-TMD and plasmonic nanoantennas show a strong dependency on the linear polarization state [24]. In this way, we determined the relative angle between the x-axis as defined in Figure 1A and the long axis of the nanorods. The results can be found in the Supplementary Material. Note that all upcoming x-axes, labeled with "Polarization angle" are related to the orientation of the individual transferred 1L-WS$_2$. Therefore, the same orientations of the antennas in the SEM images will have slightly different orientation angles with respect to the WS$_2$.

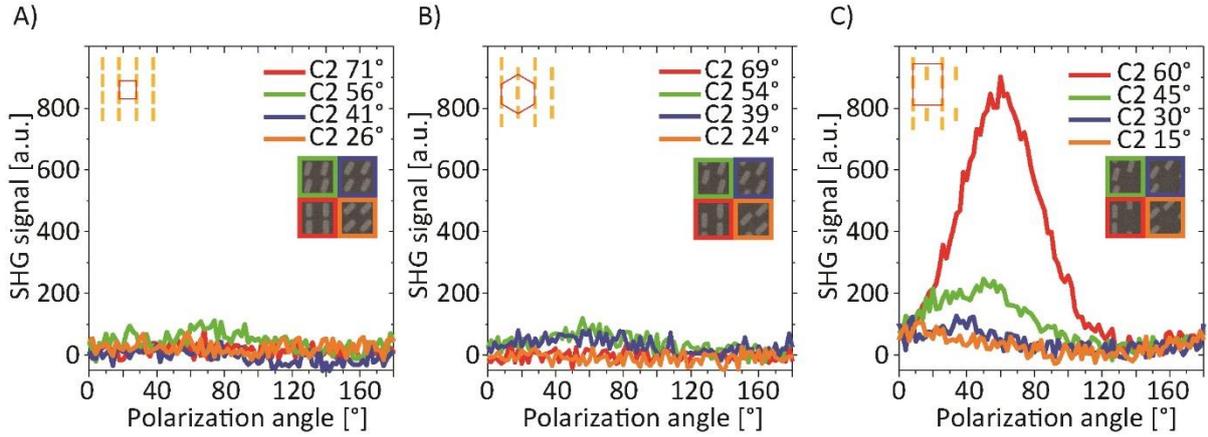

**Figure 5: Nonlinear optical measurements of plasmonic metasurfaces without the 1L-WS$_2$.** Polarization-dependent SHG signals of plasmonic metasurfaces with a square (A), hexagonal (B), and degenerated hexagonal (C) lattice structure. The upper insets show SEM images of the plasmonic nanoantenna arrays corresponding to the rotation angle of the individual nanorods labeled by the same color, while the lower insets indicate the related lattice structure of the plasmonic nanorods.

After determining the relative orientation angles, we start with the SHG of the plasmonic nanoantenna arrays *without* the 1L-WS$_2$ on top. Note that the nanorods rotation angle of each nanoantenna array is already given in respect to the x-axis, which is defined by the 1L-WS$_2$ transferred on top of the reference hybrid metasurface, which is fabricated on the same substrate. The measured SHG signals are spectrally integrated and represent the SHG signal for each polarization angle of the fundamental beam. Starting with the plasmonic metasurface with a square lattice (Figure 5A), the results show almost no SHG signal for any antenna rotation angle. Only the plasmonic nanorods, rotated by 56° in the 1L-WS$_2$ reference frame exhibit a small increase of SHG signal compared to the other rotation angles. Similar behavior is observable for the plasmonic nanorods arranged in the hexagonal lattice structure, where again, almost no SHG signal is observed (Figure 5B). As for single plasmonic nanorods with inversion symmetry, SHG is not allowed by the selection rules, which also applies to the square and hexagonal lattice structures due to their inversion symmetry. Although a small SHG signal is observable, this is more likely caused by small surface impurities, arising from the fabrication process, so that a perfect inversion symmetry is not given. In addition to a non-perfect surface, the comparatively small laser spot might give rise to a small SHG signal. When its center is not aligned perfectly with the symmetric point of the nanoantenna array's lattice structure it could lead to an imperfect cancellation of SHG from different locations. The situation changes for plasmonic nanorods arranged in the degenerated hexagonal lattice structure since the inversion symmetry is indeed broken by the lattice and SHG is allowed. Figure 5C shows the measured SHG signal, which is strongest for the antenna rotation angle of 60°. Although the other nanoantenna arrays, where the nanorods have different rotation angles, exhibit weaker SHG signals than for antennas rotated by 60°, they are still stronger compared to the SHG signals arising from the nanoantenna arrays arranged in the square and hexagonal lattice structure. The largest SHG signal occurrs when the nanorods are excited alongside its long axis due to the excitation of the localized surface plasmon resonance, which can be seen by the shifted SHG maxima for different antenna rotation angles. Since the SHG originates from the lattice structure, the coupling between the individual antennas plays an important role when it comes to SHG. The decaying SHG signal strength for greater antenna rotation angles can be explained by a weaker coupling between the individual nanorods. The coupling effect is also visible in a polarization angle dependent transmission measurement, which can be found in the Supplementary Material. The results underline, that for smaller antenna rotation angles, the transmission of the plasmonic nanoantenna arrays becomes higher, indicating a weaker excitation.

## Symmetry effects on SHG for hybrid metasurfaces

We repeated the polarization-dependent SHG measurements on the hybrid metasurfaces. Therefore, we used the same setup and parameters as for the measurements in Figure 5. For better visualization, Figures 6A-C illustrate the determined relative orientation angles between the individual nanorods and the 1L-WS$_2$ for each lattice structure for the fabricated samples.

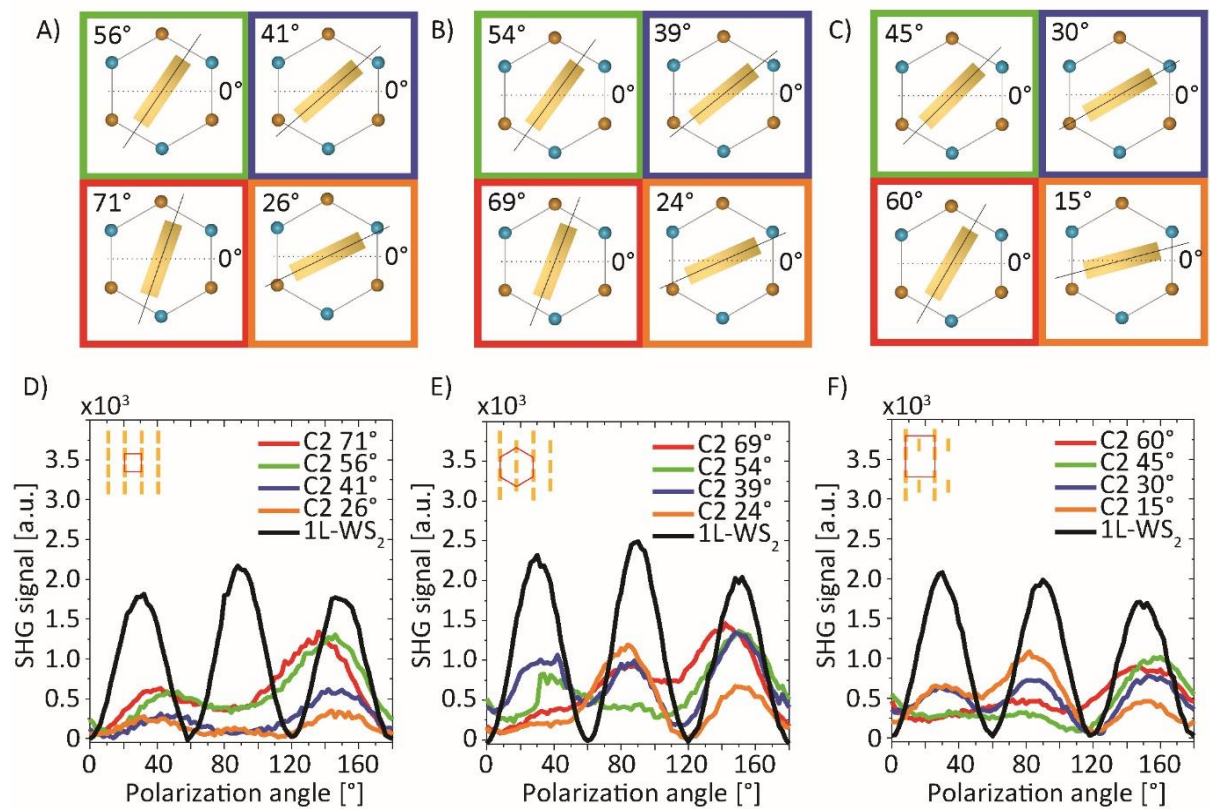

**Figure 6: SHG signals of the hybrid metasurfaces including the 1L-WS$_2$. (A-C)** Schematic illustrations of the relative orientations between the plasmonic nanorods and the hexagonal

lattice structure of the 1L-WS$_2$ for the square (A), hexagonal (B), and degenerated hexagonal (C) lattice structure. **(D-F)** Polarization-dependent SHG signals of hybrid metasurfaces with plasmonic nanoantenna arrays arranged in a square (D), hexagonal (E), and degenerated hexagonal (F) lattice structure. While the color code matches the marked colors in Figures 5(A-C), the black data represents the polarization-dependent SHG signal of bare 1L-WS$_2$. Note that the insets indicate the related lattice structure of the plasmonic nanoantenna array.

Starting with the structure with the square lattice (Figure 6D), the polarization-dependent SHG signal of the hybrid metasurface (colored) differs strongly from the SHG signal of the individual plasmonic nanoantenna arrays (Figure 5A) and 1L-WS$_2$ (black line). The strength of the SHG signal of the hybrid structure is much greater compared to the bare plasmonic metasurface case. Since the plasmonic metasurface only contributes weakly to the hybrid metasurface SHG signal, it seems to be mainly determined by the 1L-WS$_2$ and for certain polarization angles, it becomes almost as strong as for 1L-WS$_2$ alone (black). Nevertheless, the polarization dependency deviates from the bare 1L-WS$_2$ case. The SHG signal of the hybrid metasurface becomes stronger when the polarization is alongside a mirror symmetry axis of the 1L-WS$_2$, where a maximum in SHG signal strength is observable, e.g. ~150°. We find a second maximum at ~30°, albeit the SHG signal being lower than for a polarization angle of 150°. This effect can be explained by the orientation of the nanorods. The more the long axis of the nanorod is aligned with the symmetry axis of the 1L-WS$_2$, where a maximum of SHG is expected, the more the nanorod is excited by the NIR light and less amount of light participates in the SHG in the 1L-WS$_2$. The closer the long axis of the nanorod to the symmetry axis of the 1L-WS$_2$ is, the lower the SHG signal of the hybrid metasurface becomes. These observations are supported by a theoretical analysis, where we model the hybrid metasurface by splitting it into two parts, the linear transmission of the plasmonic nanorods and the SHG of the 1L-WS$_2$. A detailed discussion can be found in the supplementary material. It is shown, that the same key features, visible in the experimental data and previously discussed, are also observable in the theoretical model. Nevertheless, the experimental data show additional effects, which the theoretical model does not take into account, such as SHG in the lattice structure arising from the plasmonic nanoantenna arrays, lattice interaction between individual plasmonic nanoantennas or other possible phenomena, like dipole dipole interactions.[40]

Similar behavior is observable for the hybrid metasurface with plasmonic nanoantennas arranged in a hexagonal lattice structure, where again, the SHG signal becomes strongest for polarization angles parallel to a 1L-WS$_2$ symmetry axis and the nanorods short axis, e.g. a polarization angle of 90° and 150°. Since the inversion symmetry is still given for this lattice structure, the SHG signal strength seems also mainly determined by the 1L-WS$_2$, whereas the bare plasmonic metasurface contributes only weakly (Figure 5B). Nevertheless, the SHG signal strength of the hybrid metasurface is weaker than the one arising from 1L-WS$_2$ alone for all polarization angles. Surprisingly, the same observation can be made from the results of hybrid metasurface with plasmonic nanorods arranged in the degenerated hexagonal lattice structure for polarization angles of 90° and 150°. Although SHG is allowed for the plasmonic nanoantenna array with broken inversion symmetry, we do not observe the same SHG signal strength for the hybrid metasurface compared to the bare plasmonic case for, e.g., a polarization angle of 60°, where no additional local maximum is formed. It stands out, that the polarization-dependent SHG signals arising from all these hybrid metasurfaces are mainly determined by the 1L-WS$_2$, where its local maxima are located at 30°, 90°, and 150°, respectively. The SHG follows the polarization dependency of 1L-WS$_2$ with three distinct maxima located at nearly the same polarization angles of 30°, 90°, and 150°. However, stronger SHG signals can be detected at polarization angles of 0°, 60°, and 120°, where the bare 1L-WS$_2$ exhibits only a very weak SHG signal, which is again independent of the lattice structure of the plasmonic nanoantenna array. This hints, that the plasmonic nanorods as part of the presented hybrid metasurfaces provide a polarization channel for SHG, where it is forbidden for either 1L-WS$_2$ or bare plasmonic metasurface. To characterize the strength of these channels, we calculate the SHG increase for these cases. Therefore, we determine the enhancement $\varepsilon_{SHG}$ of the SHG by $\varepsilon_{SHG} = SHG_{hybrid}/(SHG_{PNA} + SHG_{1L-WS_2})$ with $SHG_{hybrid}$, $SHG_{PNA}$ and $SHG_{1L-WS_2}$ as the polarization-dependent absolute SHG signals, measured for the hybrid metasurfaces, plasmonic nanoantenna arrays and 1L-WS$_2$, respectively. The results are illustrated in Figures 7A-C.

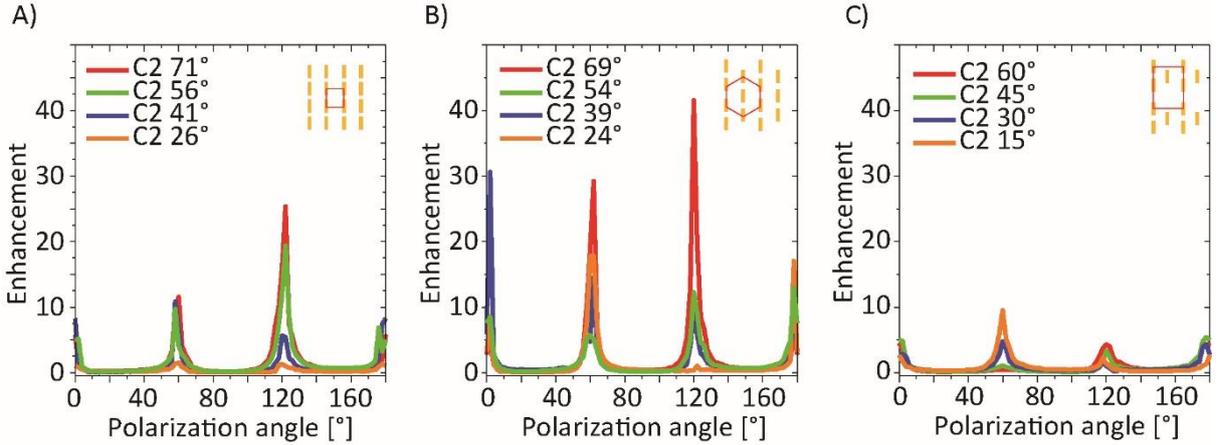

**Figure 7: Enhancement factors of SHG for hybrid metasurfaces. (A-C)** Polarization-dependent enhancement $\varepsilon_{SHG}$ for the hybrid metasurface with plasmonic nanorods arranged in a square (A), hexagonal (B), and degenerated hexagonal (C) lattice structure for different orientations of the nanoantennas. The insets in each plot indicate the related lattice structure of the plasmonic nanorod array.

The results show that the enhancement peaks at polarization angles of 0°, 60° and, 120° where no SHG signal is expected for the bare 1L-WS$_2$, meanwhile in-between these angles, no enhancement of the SHG is observable but attenuation of the SHG signal is visible. This holds true for all measured hybrid metasurfaces, independently from the lattice structure of the plasmonic nanoantenna array. This supports our previous observation, that the SHG, generated by the hybrid metasurfaces is channeled by the plasmonic nanorods into polarization states, that are forbidden for bare 1L-$WS_2$ or bare plasmonic nanoantenna arrays. By comparing the degree of enhancement among the different lattice structures, it is noticeable, that for the hybrid metasurface with the nanoantennas arranged in the degenerated hexagonal lattice structure the enhancement is lowest among all measured samples. This points out, that the SHG, generated by the plasmonic nanoantenna array does not support an overall increase of the generated SHG when it is combined with 1L-$WS_2$ for a hybrid metasurface. By comparing the two other hybrid metasurfaces with nanoantenna arrays arranged in lattice structures without broken inversion symmetry, the sample with the hexagonal lattice structure shows an almost twice as large enhancement than the sample with the square lattice structure, although both plasmonic nanoantenna arrays exhibit a weak SHG of similar strength.

As we have shown, the studied hybrid metasurfaces provide a strong enhancement of SHG in polarization channels, which are forbidden by either bare 1L-$WS_2$ and plasmonic metasurface individually, while attenuating the signal strength for other polarization channels, where usually a strong signal for bare 1L-$WS_2$ is observable. To understand, how plasmonic resonances influence the SHG in hybrid metasurfaces, we investigated three additional hybrid metasurfaces with plasmonic nanoantenna arrays, arranged in a square lattice. Different from the previous samples, we varied the length of the plasmonic nanorods to change the plasmon resonance and therefore the coupling to the 1L-$WS_2$. The samples were fabricated the same way as the previous samples and provide plasmon resonances at around 1130 nm, 1230 nm and 1380 nm for nanorod lengths of 270 nm, 300 nm and 330 nm, respectively. Corresponding transmission spectra are shown in Figure 8A where SEM images of the plasmonic nanorods are added. As the previous measurements were done with an excitation wavelength of 1230 nm, we now measure the SHG signal of the hybrid metasurfaces for excitation wavelengths ranging from 1160-1370 nm to examine, whether the enhancement changes for different excitation wavelengths and plasmon resonances. Further, we switched to a circular polarization state of the excitation beam to investigate a more generalized impact of plasmonic resonances on the SHG in hybrid metasurfaces, since the SHG of hybrid metasurfaces is strongly dependent on the linear polarization state. Therefore, we replaced the input half-wave plate from Figure 4A with a quarter-wave plate while removing the output linear polarizer. A schematic illustration of the setup used as well as further sample characterization is shown in the Supplementary Material.

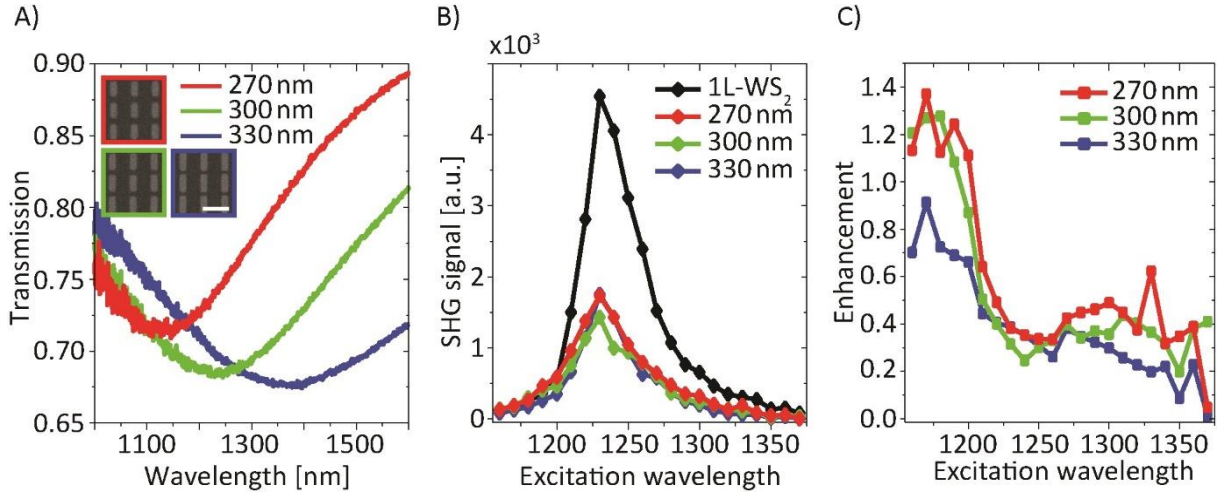

**Figure 8: Influence of the plasmonic resonance wavelength. (A)** Transmission spectra of plasmonic nanoantenna arrays with nanorod lengths of 270 nm, 300 nm, and 330 nm. The insets show SEM images of the nanorods while the white scale bar indicates 500 nm. For shorter antenna length the plasmon resonance shifts to shorter wavelengths. **(B)** SHG signals of bare 1L-WS$_2$ and the different hybrid metasurfaces for various excitation wavelengths ranging from 1160-1370 nm. **(C)** Calculated enhancement $E$ for different excitation wavelengths. Note that the SHG signal of bare plasmonic nanoantenna arrays is assumed to be zero for all excitation wavelengths since no SHG is expected.

The results of the SHG measurement for different excitation wavelengths are shown in Figure 8B. It stands out, that the SHG for all hybrid metasurfaces peaks at an excitation wavelength of 1230 nm, which is half of the band gap energy of the 1L-WS$_2$. Further, the SHG signal of the hybrid metasurfaces for this excitation wavelength is lower than for solely 1L-WS$_2$. These observations coinceide with to the previous wavelength-dependent measurement shown in Figure 4. By looking closer at the SHG signals of the hybrid metasurfaces for an excitation wavelength of 1230 nm, the hybrid metasurface with nanorods of 300 nm length exhibits a lower SHG than the other two hybrid metasurfaces. This can be explained by a more resonant excitation of the nanorods, resulting in an attenuated SHG in hybrid metasurfaces. This effect was already observed in our previous polarization-dependent measurements, shown in Figure 6. To further investigate the impact of the plasmon resonances on the SHG in hybrid metasurfaces, we now calculate the enhancement $\varepsilon_{\text{SHG}}$ for each excitation wavelength again. Note, that we neglect the SHG signal for the bare plasmonic metasurface since no SHG signal is expected for plasmonic nanoantenna arrays arranged in a square lattice, as we have already demonstrated in Figure 5A. The results of the calculated enhancement for the different hybrid metasurfaces are shown in Figure 8C. It stands out, that for all the different antenna lengths, the SHG is attenuated when excited with wavelengths longer than 1200 nm. Further, the SHG signal of the hybrid metasurfaces stays comparatively low to the SHG signal of pure 1L-WS$_2$ for all subsequent wavelengths, while the hybrid metasurface with nanorods of 300 nm length attenuate the SHG at 1230 nm the most.

By looking at shorter wavelengths than 1200 nm, a small enhancement of SHG is observable for the hybrid metasurfaces with nanorod lengths of 270 nm and 300 nm. Although this is not the case for the hybrid metasurface with nanorods length of 330 nm, a comparable increase of the enhancement is observable. This leads to the conclusion, that the SHG can be enhanced when hybrid metasurfaces are excited with photon energies, higher than half of the band bap energy of the 1L-WS$_2$. At the same time, plasmon resonances can support this enhancement, since the results in Figure 8C show, that plasmon resonances at shorter wavelengths support a greater enhancement than plasmon resonances at higher wavelengths.

## Conclusion

We presented experimental results of the polarization-dependent SHG in hybrid metasurfaces consisting of plasmonic nanoantenna arrays arranged in different lattice structures covered with a monolayer of WS$_2$. The linear optical characterization of the individual components showed a good overlap between the resonances of the plasmonic nanoantenna arrays with half of the band gap energy of the 1L-WS$_2$. We characterized the nonlinear optical response from both parts of the hybrid metasurface for different linear input polarization states to investigate differences from the nonlinear optical response of the hybrid metasurface. These measurements showed a strong impact of the plasmonic nanorods on the polarization-dependent SHG, where we measured a stronger SHG signal when the nanorods are not excited resonantly. It was shown, that measuring the SHG signal at operating wavelengths at half of the band gap energy of the 1L-WS$_2$, its nonlinearity is already strong and the SHG is not further enhanced by plasmon resonances. From the measurements we conclude that plasmonic nanoantennas not always will lead to an increased nonlinear signal due to their field enhancement at their surface. The overall loss introduced by the metal might counteract the enhancement and result in an overall decrease of the nonlinear response.

However, we found that the plasmonic nanorods rather provide a channel for SHG in polarization states, which are usually forbidden by either 1L-WS$_2$ or plasmonic nanoantennas alone. To quantify the potential of these channels, we calculated the enhancement of SHG for the different polarization angles and determined the enhancement of SHG in these channels to a factor of more than 40. This enhancement is strongly affected by the lattice structure of the plasmonic nanoantenna array and becomes lowest for hybrid metasurfaces with plasmonic nanoantenna arrays arranged in a degenerated hexagonal lattice structure with broken inversion symmetry, while it increases for hybrid metasurfaces with plasmonic nanoantenna arrays without broken inversion symmetry. This behavior can be explained by an already strong SHG signal, that a bare plasmonic metasurface with a broken inversion symmetry provides, leading to a lower enhancement factor. As we further analyzed the enhancement, we found that it is not only affected by certain polarization states, but also by the excitation wavelength and the plasmon resonances. The excitation of a hybrid metasurface with wavelengths at half of the band gap energy only leads to an enhanced SHG into polarization channels forbidden by either 1L-$WS_2$ or bare plasmonic nanoantenna array. This changes, when the hybrid metasurface is excited at wavelengths with photon energies lower than half of the band gap energy. Here, we detect a small enhancement of SHG for circular polarization states, which are supported by plasmon resonances located nearby.

In conclusion, we studied the process of second harmonic generation in hybrid metasurfaces for different orientations of the plasmonic antennas with respect to the hexagonal lattice structure of the $WS_2$ and provide experimental results on how it is affected by certain symmetries of its individual components as well as excitation wavelengths and plasmon resonances. By operating at wavelengths, lower than half of the band gap energy of 1L-$WS_2$, an enhanced SHG can be measured and might be further increased, if certain polarization states are chosen.  As these results provide an insight on the interaction of plasmonic nanoantennas and 1L-$WS_2$ for nonlinear processes, further investigations of hybrid metasurfaces with plasmonic nanoantenna arrays containing nanoantennas of different symmetries than the C2 rotational symmetry might provide more degrees of freedom for tailoring nonlinear properties. As it was shown, different symmetries in hybrid metasurfaces provide a strong SHG enhancement, new ways may arise to further support the SHG with different antenna geometries for various polarization states.

## Methods

First, we fabricate the metasurfaces by standard electron beam lithography. After the lithography is done on a resist spin-coated fused quartz substrate, the resist mask is developed and the sample is coated with a 30 nm thick gold layer by electron beam evaporation. In the subsequent step, the mask is removed in a lift-off process so that the nanorods remain on the substrate surface. After the linear optical characterization of the fabricated plasmonic nanoantenna arrays, we transferred the 1L-$WS_2$ from a PDMS (polydimethylsiloxane) stamp.

## Acknowledgments

The authors acknowledge the funding provided by the European Research Council (ERC) under the European Union's Horizon 2020 research and innovation program (grant agreement No. 724306) and the Deutsche Forschungsgemeinschaft (DFG, German Research Foundation)–SFB-Geschäftszeichen TRR142/2-2020–Projektnummer 231447078–Teilprojekt A07/A08.

## Supporting information

The supporting information contains further details to the fabricated plasmonic nanoantenna arrays as well as PL-images of the investigated hybrid metasurfaces. In addition, nonlinear measurements are presented supporting the discussion in this work.

## References

1. Two-Dimensional Transition-Metal Dichalcogenides Alexander V. Kolobov and Junji Tominaga: Springer. **2016**.
2. Mak, K. F.; Lee, C.; Hone, J.; Shan, J.; Heinz, T. F., Atomically Thin MoS2: A New Direct-Gap Semiconductor. *Phys Rev Lett* **2010**, *105* (13), 136805.
3. Xie, L. M., Two-dimensional transition metal dichalcogenide alloys: preparation, characterization and applications. *Nanoscale* **2015**, *7* (44), 18392-18401.
4. Li, Y. L.; Chernikov, A.; Zhang, X.; Rigosi, A.; Hill, H. M.; van der Zande, A. M.; Chenet, D. A.; Shih, E. M.; Hone, J.; Heinz, T. F., Measurement of the optical dielectric function of monolayer transition-metal dichalcogenides: MoS2, MoSe2, WS2, and WSe2. *Phys Rev B* **2014**, *90* (20), 205422.
5. Zhao, W.; Ghorannevis, Z.; Chu, L.; Toh, M.; Kloc, C.; Tan, P.-H.; Eda, G., Evolution of Electronic Structure in Atomically Thin Sheets of WS2 and WSe2. *Acs Nano* **2013**, *7* (1), 791-797.
6. Gutiérrez, H. R.; Perea-López, N.; Elías, A. L.; Berkdemir, A.; Wang, B.; Lv, R.; López-Urías, F.; Crespi, V. H.; Terrones, H.; Terrones, M., Extraordinary Room-Temperature Photoluminescence in Triangular WS2 Monolayers. *Nano Lett* **2013**, *13* (8), 3447-3454.
7. Eda, G.; Yamaguchi, H.; Voiry, D.; Fujita, T.; Chen, M.; Chhowalla, M., Photoluminescence from Chemically Exfoliated MoS2. *Nano Lett* **2011**, *11* (12), 5111-5116.
8. Lee, C.; Yan, H.; Brus, L. E.; Heinz, T. F.; Hone, J.; Ryu, S., Anomalous Lattice Vibrations of Single- and Few-Layer MoS2. *Acs Nano* **2010**, *4* (5), 2695-2700.
9. Wang, H.; Zhang, C.; Chan, W.; Tiwari, S.; Rana, F., Ultrafast response of monolayer molybdenum disulfide photodetectors. *Nat Commun* **2015**, *6* (1), 8831.
10. Yin, Z.; Li, H.; Li, H.; Jiang, L.; Shi, Y.; Sun, Y.; Lu, G.; Zhang, Q.; Chen, X.; Zhang, H., Single-Layer MoS2 Phototransistors. *Acs Nano* **2012**, *6* (1), 74-80.
11. Mak, K. F.; Shan, J., Photonics and optoelectronics of 2D semiconductor transition metal dichalcogenides. *Nat Photonics* **2016**, *10* (4), 216-226.
12. Liu, W. J.; Lee, B.; Naylor, C. H.; Ee, H. S.; Park, J.; Johnson, A. T. C.; Agarwal, R., Strong Exciton-Plasmon Coupling in MoS2 Coupled with Plasmonic Lattice. *Nano Lett* **2016**, *16* (2), 1262-1269.
13. Kleemann, M. E.; Chikkaraddy, R.; Alexeev, E. M.; Kos, D.; Carnegie, C.; Deacon, W.; de Pury, A. C.; Grosse, C.; de Nijs, B.; Mertens, J.; Tartakovskii, A. I.; Baumberg, J. J., Strong-coupling of WSe2 in ultra-compact plasmonic nanocavities at room temperature. *Nat Commun* **2017**, *8*, 1296.
14. Singh, M. R.; Najiminaini, M.; Balakrishnan, S.; Carson, J. J. L., Metamaterial-based theoretical description of light scattering by metallic nano-hole array structures. *Journal of Applied Physics* **2015**, *117* (18), 184302.
15. Singh, M. R., Enhancement of the second-harmonic generation in a quantum dot–metallic nanoparticle hybrid system. *Nanotechnology* **2013**, *24* (12), 125701.

16. Mundry, J.; Spreyer, F.; Jmerik, V.; Ivanov, S.; Zentgraf, T.; Betz, M., Nonlinear metasurface combining telecom-range intersubband transitions in GaN/AlN quantum wells with resonant plasmonic antenna arrays. *Opt. Mater. Express* **2021,** *11* (7), 2134-2144.

17. Cong, C. X.; Shang, J. Z.; Wu, X.; Cao, B. C.; Peimyoo, N.; Qiu, C.; Sun, L. T.; Yu, T., Synthesis and Optical Properties of Large-Area Single-Crystalline 2D Semiconductor WS2 Monolayer from Chemical Vapor Deposition. *Adv Opt Mater* **2014,** *2* (2), 131-136.

18. Liu, F.; Wu, W.; Bai, Y.; Chae, S. H.; Li, Q.; Wang, J.; Hone, J.; Zhu, X.-Y., Disassembling 2D van der Waals crystals into macroscopic monolayers and reassembling into artificial lattices. *Science* **2020,** *367* (6480), 903-906.

19. He, T.; Li, Y.; Zhou, Z.; Zeng, C.; Qiao, L.; Lan, C.; Yin, Y.; Li, C.; Liu, Y., Synthesis of large-area uniform MoS 2 films by substrate-moving atmospheric pressure chemical vapor deposition: from monolayer to multilayer. *2d Mater* **2019,** *6* (2), 025030.

20. Gong, Y.; Ye, G.; Lei, S.; Shi, G.; He, Y.; Lin, J.; Zhang, X.; Vajtai, R.; Pantelides, S. T.; Zhou, W.; Li, B.; Ajayan, P. M., Synthesis of Millimeter-Scale Transition Metal Dichalcogenides Single Crystals. *Advanced Functional Materials* **2016,** *26* (12), 2009-2015.

21. Wang, S. J.; Li, S. L.; Chervy, T.; Shalabney, A.; Azzini, S.; Orgiu, E.; Hutchison, J. A.; Genet, C.; Samori, P.; Ebbesen, T. W., Coherent Coupling of WS2 Monolayers with Metallic Photonic Nanostructures at Room Temperature. *Nano Lett* **2016,** *16* (7), 4368-4374.

22. Chen, X. X.; Wang, H.; Xu, N. S.; Chen, H. J.; Deng, S. Z., Resonance coupling in hybrid gold nanohole-monolayer WS2 nanostructures. *Appl Mater Today* **2019,** *15*, 145-152.

23. Mukherjee, B.; Kaushik, N.; Tripathi, R. P. N.; Joseph, A. M.; Mohapatra, P. K.; Dhar, S.; Singh, B. P.; Kumar, G. V. P.; Simsek, E.; Lodha, S., Exciton Emission Intensity Modulation of Monolayer MoS2 via Au Plasmon Coupling. *Sci Rep-Uk* **2017,** *7*, 41175.

24. Malard, L. M.; Alencar, T. V.; Barboza, A. P. M.; Mak, K. F.; de Paula, A. M., Observation of intense second harmonic generation from MoS2 atomic crystals. *Phys Rev B* **2013,** *87* (20), 201401.

25. Kumar, N.; Najmaei, S.; Cui, Q.; Ceballos, F.; Ajayan, P. M.; Lou, J.; Zhao, H., Second harmonic microscopy of monolayer MoS2. *Phys Rev B* **2013,** *87* (16), 161403.

26. Zeng, H.; Liu, G.-B.; Dai, J.; Yan, Y.; Zhu, B.; He, R.; Xie, L.; Xu, S.; Chen, X.; Yao, W.; Cui, X., Optical signature of symmetry variations and spin-valley coupling in atomically thin tungsten dichalcogenides. *Sci Rep-Uk* **2013,** *3* (1), 1608.

27. Hu, G. W.; Hong, X. M.; Wang, K.; Wu, J.; Xu, H. X.; Zhao, W. C.; Liu, W. W.; Zhang, S.; Garcia-Vidal, F.; Wang, B.; Lu, P. X.; Qiu, C. W., Coherent steering of nonlinear chiral valley photons with a synthetic Au-WS2 metasurface. *Nat Photonics* **2019,** *13* (7), 467–472.

28. Chen, J. W.; Wang, K.; Long, H.; Han, X. B.; Hu, H. B.; Liu, W. W.; Wang, B.; Lu, P. X., Tungsten Disulfide-Gold Nanohole Hybrid Metasurfaces for Nonlinear Metalenses in the Visible Region. *Nano Lett* **2018,** *18* (2), 1344-1350.

29. Spreyer, F.; Zhao, R.; Huang, L.; Zentgraf, T., Second harmonic imaging of plasmonic Pancharatnam-Berry phase metasurfaces coupled to monolayers of WS2. *Nanophotonics* **2020,** *9* (2), 351-360.

30. Han, X.; Wang, K.; Persaud, P. D.; Xing, X.; Liu, W.; Long, H.; Li, F.; Wang, B.; Singh, M. R.; Lu, P., Harmonic Resonance Enhanced Second-Harmonic Generation in the Monolayer WS2–Ag Nanocavity. *Acs Photonics* **2020,** *7* (3), 562-568.

31. Li, G. X.; Chen, S. M.; Pholchai, N.; Reineke, B.; Wong, P. W. H.; Pun, E. Y. B.; Cheah, K. W.; Zentgraf, T.; Zhang, S., Continuous control of the nonlinearity phase for harmonic generations. *Nat Mater* **2015,** *14* (6), 607-612.

32. Wang, Z.; Dong, Z. G.; Zhu, H.; Jin, L.; Chiu, M. H.; Li, L. J.; Xu, Q. H.; Eda, G.; Maier, S. A.; Wee, A. T. S.; Qiu, C. W.; Yang, J. K. W., Selectively Plasmon-Enhanced Second-Harmonic Generation from Monolayer Tungsten Diselenide on Flexible Substrates. *Acs Nano* **2018,** *12* (2), 1859-1867.
33. Shi, J. W.; Liang, W. Y.; Raja, S. S.; Sang, Y. G.; Zhang, X. Q.; Chen, C. A.; Wang, Y. R.; Yang, X. Y.; Lee, Y. H.; Ahn, H.; Gwo, S., Plasmonic Enhancement and Manipulation of Optical Nonlinearity in Monolayer Tungsten Disulfide. *Laser Photonics Rev* **2018,** *12* (10), 1800188.
34. Li, Y.; Kang, M.; Shi, J.; Wu, K.; Zhang, S.; Xu, H., Transversely Divergent Second Harmonic Generation by Surface Plasmon Polaritons on Single Metallic Nanowires. *Nano Lett* **2017,** *17* (12), 7803-7808.
35. Wang, X.; Gogol, P.; Cambril, E.; Palpant, B., Near- and Far-Field Effects on the Plasmon Coupling in Gold Nanoparticle Arrays. *The Journal of Physical Chemistry C* **2012,** *116* (46), 24741-24747.
36. Xu, T.; Jiao, X.; Zhang, G. P.; Blair, S., Second-harmonic emission from sub-wavelength apertures: Effects of aperture symmetry and lattice arrangement. *Opt. Express* **2007,** *15* (21), 13894-13906.
37. Humphrey, A. D.; Barnes, W. L., Plasmonic surface lattice resonances on arrays of different lattice symmetry. *Phys Rev B* **2014,** *90* (7), 075404.
38. Xiao, J.; Ye, Z. L.; Wang, Y.; Zhu, H. Y.; Wang, Y.; Zhang, X., Nonlinear optical selection rule based on valley-exciton locking in monolayer $WS_2$. *Light-Sci Appl* **2015,** *4*, e366.
39. Wang, G.; Marie, X.; Gerber, I.; Amand, T.; Lagarde, D.; Bouet, L.; Vidal, M.; Balocchi, A.; Urbaszek, B., Giant Enhancement of the Optical Second-Harmonic Emission of WSe2 Monolayers by Laser Excitation at Exciton Resonances. *Phys Rev Lett* **2015,** *114* (9), 097403.
40. Singh, M. R.; Persaud, P. D., Dipole–Dipole Interaction in Two-Photon Spectroscopy of Metallic Nanohybrids. *The Journal of Physical Chemistry C* **2020,** *124* (11), 6311-6320.

# Supplementary Material

**Plasmonic nanoantenna arrays**

The plasmonic nanoantenna arrays, as part of the hybrid metasurfaces, investigated in this work, consist of nanorods with a C2 rotational symmetry. The individual nanorods are fabricated by standard electron beam lithography on a quartz substrate with a length of 300 nm, width of 100 nm. For lithography, the substrate is spin-coated with a poly methyl methacrylate (PMMA) resist. After lithography, the PMMA resist is developed and the sample is placed into an electron beam evaporator for gold deposition of 30 nm. Subsequently, the sample is placed into acetone for lift-off process, where the pmma is removed while the gold nanorods remain on the surface as the plasmonic nanoantenna array. In this work, we fabricated three different samples with C2 antennas arranged in three different lattice structures, named as square, hexagonal and degenerated hexagonal lattice, depicted in Figure 1S A-C, respectively.

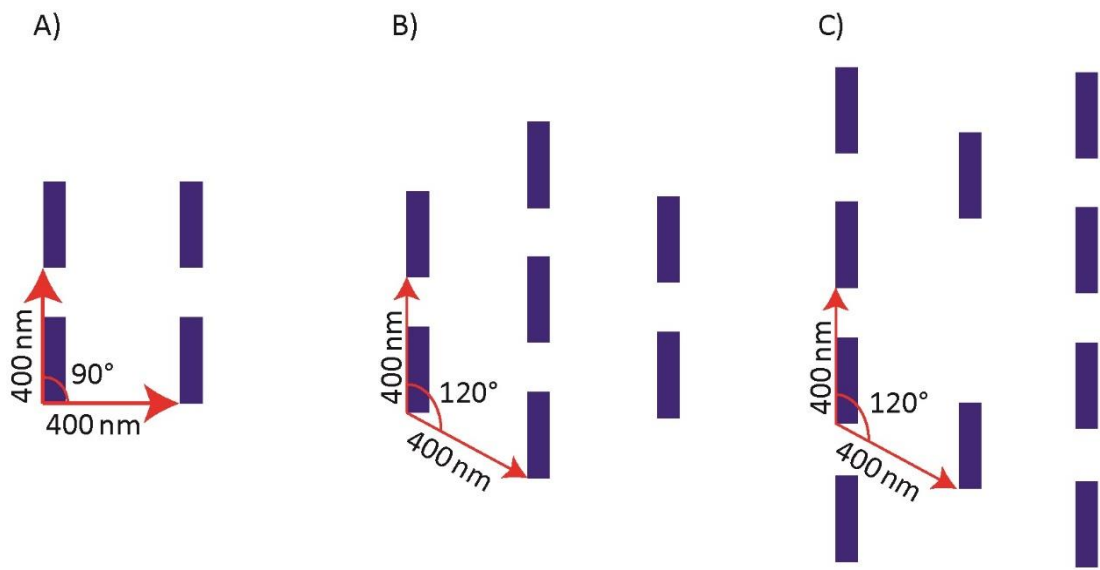

**Figure S9: Dimensions of plasmonic nanoantenna arrays** arranged in a **(A)** square, **(B)** hexagonal and **(C)** degenerated hexagonal lattice structure. The lattice constant is fixed to 400 nm, while the individual nanorods are 300 nm in length, 100 nm in width and 30 nm in height.

**Photoluminescence measurements**

In order to determine whether the transferred $WS_2$ is of monolayer thickness, we measure the photoluminescence (PL) of it, by exciting the layer with a laser beam of 532 nm, emitted by a laser diode. The beam is focused by a lens onto the sample, where the $WS_2$ covers the plasmonic nanoantenna array. The emitted PL light is collected by a microscope objective (x20/NA 0.4) and filtered by a bandpass filter (BP), so that the excitation wavelengths are blocked. Subsequently the PL light is split by a beam splitter and is detected by a camera and a spectrometer. A schematic illustration of the setup is shown in Figure S2A. The resulting PL spectra are shown in the main text in Figure 3D, while images of the 1L-WS$_2$ monolayers emitted PL light are shown in Figure S2B-D for the 1L-WS$_2$ transferred on the plasmonic nanoantenna arrays arranged in square, hexagonal and degenerated hexagonal lattices, respectively.

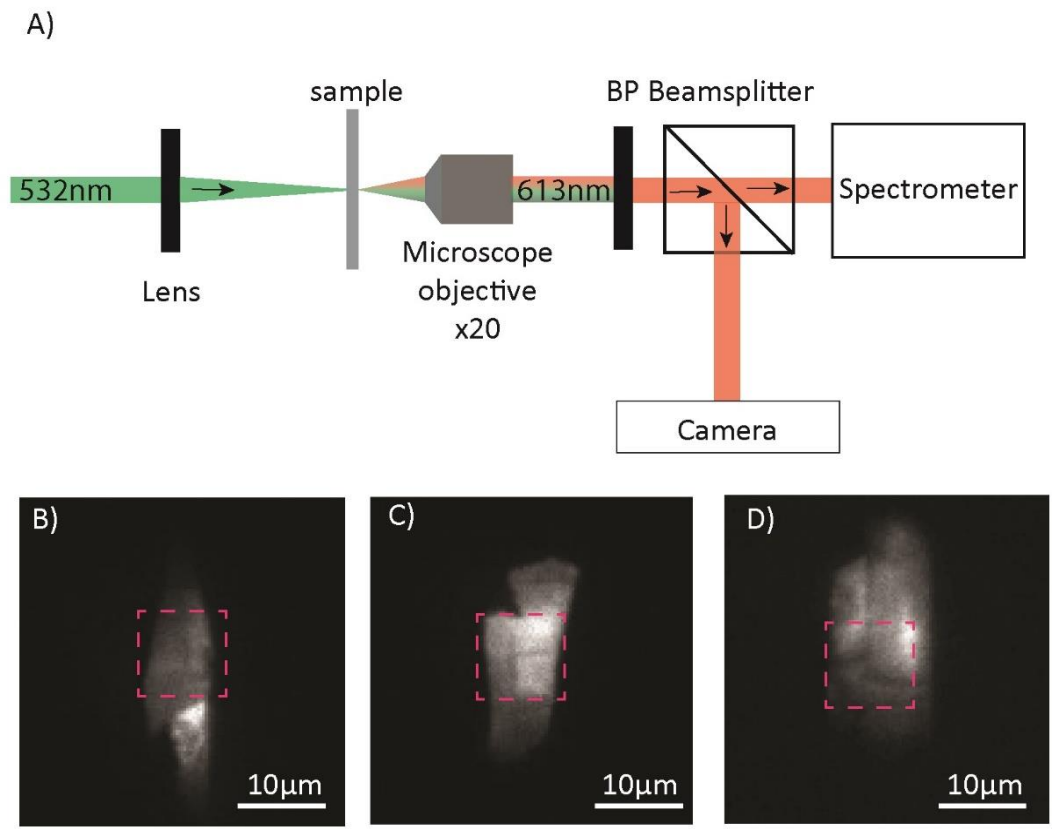

**Figure S10: Photoluminescence measurements (A)** Setup for PL-measurement. PL-images of WS$_2$ monolayer on top of the plasmonic nanoantenna arrays with **(B)** square, **(C)** hexagonal and **(D)** degenerated hexagonal lattice structures. The red dotted squares in each figure indicates the position of the plasmonic nanoantenna arrays.

## Relative angles between WS$_2$ and the nanorod antennas

In order to determine the relative orientation angles between the C2 antennas long axis and the x-axis, as defined in Figure 1A in the main text, we measure the nonlinear signals of the plasmonic nanoantenna arrays and the 1L-WS$_2$. Since both systems, individually show a strong dependence of the nonlinear signal on the excitation polarization, we rotate linear input polarization from 0-180° in 4° steps and measure the nonlinear signal, which is parallel polarized to the input polarization. The setup used is shown in Figure 4A in the main text. As for 1L-WS$_2$ and the plasmonic nanoantenna array with a degenerated hexagonal lattice structure, the nonlinear signal is generated by a process of second order, the second harmonic generation (SHG). Since SHG is only allowed for systems with broken inversion symmetry, this method is not practical for the plasmonic nanoantenna arrays arranged with square or hexagonal lattice structures due to the lack of broken inversion symmetry. For these nanoantenna arrays, we do not measure the SHG but the third harmonic generation (THG). The THG is emitted by the individual nanorods and shows the same polarization dependency on the linear excitation polarization similar to the nanorods arranged in the degenerated hexagonal lattice structure. After measuring the polarization dependent nonlinear signal for each individual component of the different hybrid metasurfaces, we calculate the particular orientation $\theta_C$ angles by fitting the function $A \cdot \sin(\pi \frac{\theta - \theta_C}{w}) + A_0$. The resulting relative orientations are given in Table ST3:

| Lattice structure | $\theta_{C_{C2}}$ | $\theta_{C_{WS_2}}$ | $\theta_{C_{C2}} - \theta_{C_{WS_2}}$ |
|---|---|---|---|
| Square | 73° | 2° | 71° |
| Hexagonal | 76° | 7° | 69° |
| Degenerated hexagonal | 82° | 22° | 60° |

**Table ST3: Orientation angles** of 1L-WS$_2$ $\left(\theta_{C_{C2}}\right)$ and plasmonic C2 nanoantennas $\left(\theta_{C_{WS_2}}\right)$ determined from polarization dependent measurement, shown in Figure S3.

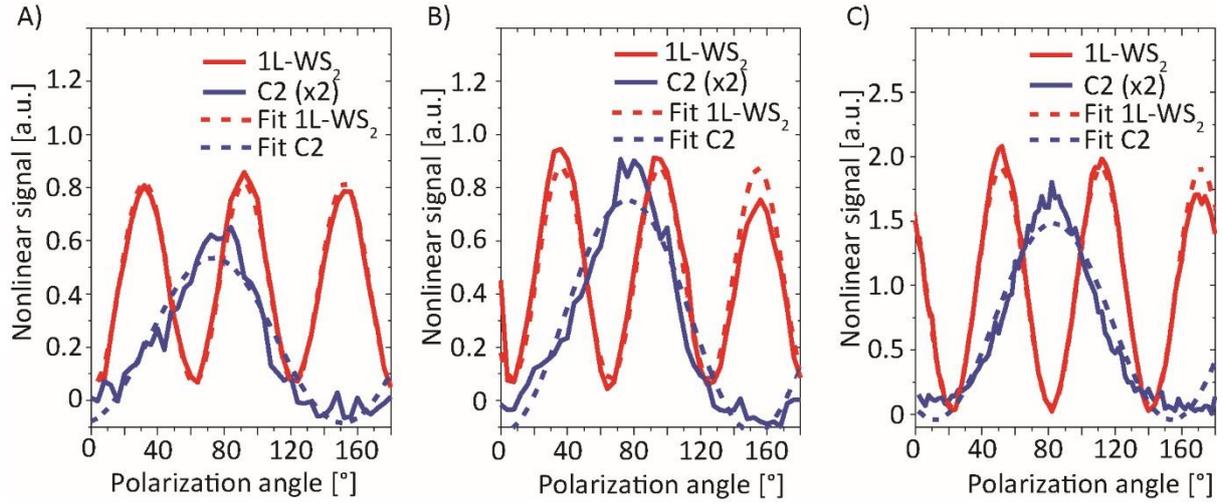

**Figure S11: Determination of relative angles between C2 antenna and WS$_2$.** **(A)** Polarization dependent nonlinear signal of 1L-WS$_2$ and C2 nanoantenna arrays arranged in a square lattice in dependency of the polarization angle. The rotation angle of the 1L-WS$_2$ is determined to 2° while the rotation angle of the C2 antenna is determined to 73°, which results in a relative rotation angle between both systems of 71°. **(B)** Polarization dependent nonlinear signal of 1L-WS$_2$ and C2 nanoantenna arrays arranged in a hexagonal lattice in dependency of the polarization angle. The rotation angle of the 1L-WS$_2$ is determined to 7° while the rotation angle of the C2 antenna is determined to 76°, which results in a relative rotation angle between both systems of 69°. **(C)** SHG signal of 1L-WS$_2$ and C2 nanoantenna arrays arranged in a degenerated hexagonal lattice in dependency of the polarization angle. The rotation angle of the 1L-WS$_2$ is determined to 22° while the rotation angle of the C2 antenna is determined to 82°, which results in a relative rotation angle between both systems of 60°. Note that all data, labeled with (x2) is multiplied with a factor of 2 for better visibility.

## Polarization dependent linear transmission

Here, we examine the coupling between the individual nanorods arranged in a degenerated hexagonal lattice structure for different rotation angles of the individual nanorods. Therefore, we used the same setup, illustrated in Figure 4A in the main text, but removed the output shortpass filter and linear polarizer. As before, we rotated the input polarization and measured the transmitted IR light with a spectrometer. Subsequently, the measured spectrum is normalized to the input spectrum in order to receive the transmission of the plasmonic nanorods and the transmission value at 1230 nm for each polarization angle is extracted. The results are shown in Figure S4. It is shown, that for different rotation angles of the individual nanorods, the transmission changes significantly. First of all, most of the light is absorbed, when the nanorods are excited resonantly along their long axis under 15°, 30°, 45° and 60°, respectively. In addition, for higher rotation angles, the coupling among the nanorods increases, resulting in a lower transmission hinting towards a higher absorption, which supports a greater potential for SHG.

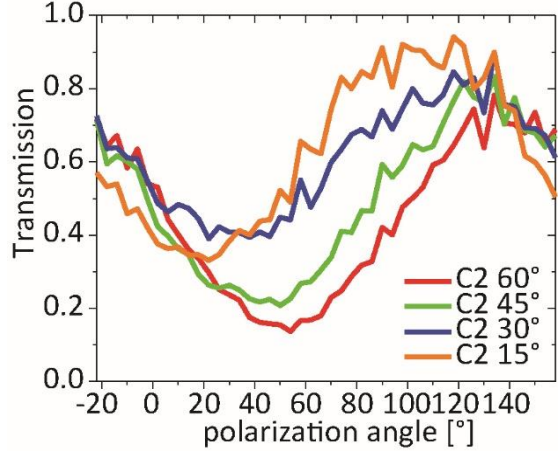

**Figure S12: Polarization dependent transmission** for 1230 nm excitation wavelength measured from the plasmonic nanoantenna arrays arranged in a degenerated hexagonal lattice structure.

## Analytical model for SHG in hybrid metasurfaces

Here, we model a theoretical approach to calculate the SHG in hybrid metasurfaces consisting of plasmonic nanoantenna arrays and 1L-WS$_2$. In our model, we split the hybrid metasurface into two parts, the linear response of the plasmonic nanorods and nonlinear second harmonic generation in monolayer WS$_2$. First, we calculate the transmitted electric field after it passed the plasmonic nanorods. Analogue to the experiment, we consider a linearly polarized electric field at normal incidence (z-direction) rotated by the angle $\theta$, which is given by

$$\vec{E} = E_0 \begin{pmatrix} \cos\theta \\ \sin\theta \end{pmatrix}$$

The Jones matrix of the plasmonic nanorods, which are rotated by an angle $\alpha$ is given by

$$J(\alpha) = \begin{bmatrix} \cos\alpha & -\sin\alpha \\ \sin\alpha & \cos\alpha \end{bmatrix} \begin{bmatrix} t_x & 0 \\ 0 & t_y \end{bmatrix} \begin{bmatrix} \cos\alpha & \sin\alpha \\ -\sin\alpha & \cos\alpha \end{bmatrix}$$

And can be used to calculate the transmitted electric field:

$$\vec{E}_{trans} = J(\alpha)\,\vec{E}$$
$$= E_0 \left[\cos\alpha \sin\alpha\,(t_x - t_y)\begin{pmatrix}\sin\theta\\\cos\theta\end{pmatrix}\right.$$
$$\left.+ \begin{pmatrix}\cos\theta\,(t_x \cos^2\alpha + t_y \sin^2\alpha)\\\sin\theta\,(t_y \cos^2\alpha + t_x \sin^2\alpha)\end{pmatrix}\right].$$

Note, that due to the thickness of the plasmonic nanorods, a phase shift $\phi$ between the two electric field components needs to be taken into account. Therefore, the two transmission coefficients of the plasmonic nanorods $t_x$ and $t_y$ are chosen to $t_x = 0.5$ (based on experimental results from Figure S4) and $t_y = e^{i\phi}$ with $\phi = 0.24\pi$ for a nanoantenna thickness of 30 nm. In the next step, we consider the SHG process in the monolayer. Since the 1L-WS$_2$ belongs to the $D_{3h}$ point symmetry group, the non-zero elements of the second order susceptibility tensor $\chi^{(2)}$ are

$$\chi^{(2)}_{yyy} = -\chi^{(2)}_{yxx} = -\chi^{(2)}_{xxy} = -\chi^{(2)}_{xyx},$$

where $x$ and $y$ denote the zigzag and armchair direction, respectively. As the second order polarization corresponds to

$$\vec{P}^{(2)} = \varepsilon_0 \chi^{(2)} \vec{E}^2_{trans} = \varepsilon_0 \chi^{(2)}_{yyy} \cdot \begin{pmatrix}-2\cdot E_{trans,x}\cdot E_{trans,y}\\E^2_{trans,y} - E^2_{trans,x}\end{pmatrix},$$

the SHG signal, parallel to the incident polarization can be calculated by

$$I^{SHG}_{\parallel} = \frac{c_0}{2\cdot\varepsilon_0}\cdot\left|\vec{P}^{(2)}\cdot\begin{pmatrix}\cos\theta\\\sin\theta\end{pmatrix}\right|^2.$$

With the help of this result, one can calculate the expected polarization dependent SHG signals, which is illustrated by Figure S5 for the different antenna orientations, and described in the main text.

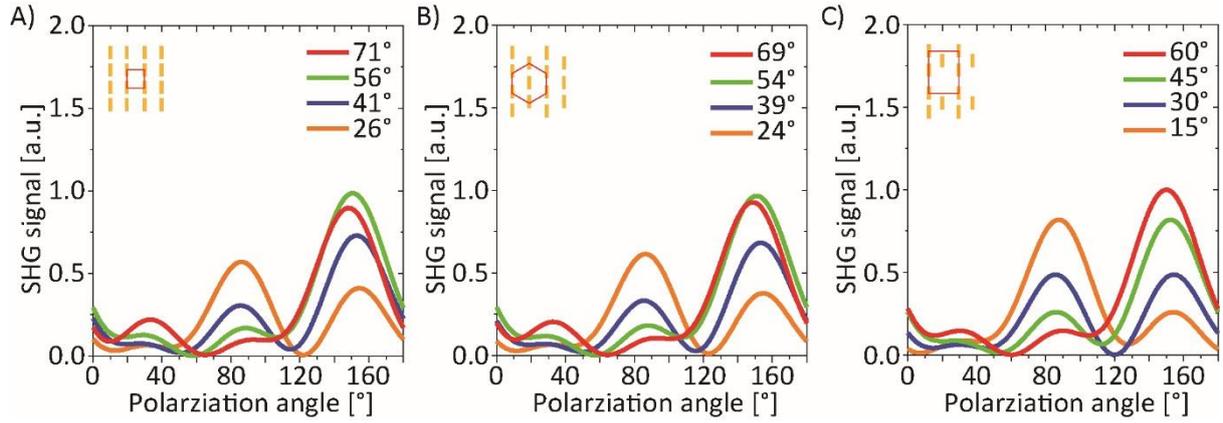

**Figure S13: SHG simulation of hybrid metasurfaces. (A-C)** Calculated polarization-dependent SHG signals of hybrid metasurfaces with plasmonic nanoantennas rotated by the denoted rotation angles $\alpha$ for the square (A), hexagonal (B), and degenerated hexagonal (C) lattice structure. Note, that the color code matches the marked colors in Figures D(D-F) in the main text and that the absolute values are not comparable to the values, shown in the main text.

As Figure S5 shows, the predominant three local maxima, originating from the hexagonal lattice structure of the 1L-$WS_2$, are still visible at polarization angles of 30°, 90° and 150°, although they underlie small shifts, which are also visible in the experimental data (Figure 6 in the main text). In addition, the calculated data shows a stronger SHG signal, when the polarization is alongside a mirror axis of the 1L-$WS_2$ while the nanoantenna is not excited resonantly along its long axis. If the nanoantenna is excited more resonantly, the SHG strength at polarization channels, where usually a strong SHG signal is expected, is attenuated. These observations support the experimental gained data, shown in the main text. Note, that our model does not consider any lattice interaction effects between the individual nanorods nor the possible process of SHG in the lattice structure of the plasmonic nanoantenna arrays or any further coupling between the nanorods and the 1L-$WS_2$.

**Wavelength dependent SHG measurements**

To determine the impact of plasmon resonances on the SHG in hybrid metasurfaces, three additional plasmonic nanoantennas were fabricated with lengths of 270 nm, 300 nm, and 330 nm. Afterwards, a monolayer of $WS_2$ was transferred onto these nanoantenna arrays. To confirm monolayer thickness, a PL measurement was done and the results are shown in Figure S5A. The PL spectrum confirms the monolayer thickness of the WS2 with a peak at 613 nm, similar to the previous measured sample, shown in Figure 3 in the main text. Subsequently, the wavelength-dependent SHG of the hybrid metasurface was measured. Therefore, the setup illustrated in Figure S5B was used. First, the fundamental beam passes a linear polarizer and a quarter-wave plate for a circular polarization. The beam then passes a long pass filter, and is focused by a microscope objective (x50/NA 0.42) onto the backside of the sample. The SHG signal is collected by another microscope objective (x100/NA 0.8), filtered by a short pass filter (SP), and measured by a spectrometer.

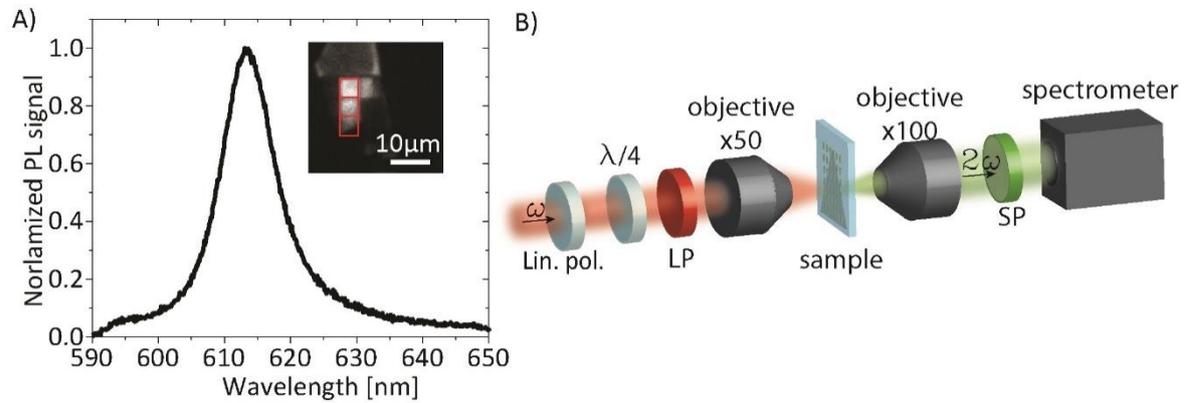

**Figure S14: PL measurement and SHG setup. (A)** Measured PL spectrum of the hybrid metasurface sample with plasmonic nanorods of length 270 nm, 300 nm, and 330 nm. The inset shows a PL image of the hybrid metasurfaces, marked in red. **(B)** Used setup for a wavelength-dependent SHG analysis.